\documentstyle[11pt,paspconf,epsf]{article}

\begin{document}

\title{The luminosity function and clustering evolution of Quasars.}
\author{S. Cristiani\altaffilmark{1},
        F. La Franca\altaffilmark{2} and 
        P. Andreani\altaffilmark{1}
}
\altaffiltext{1}
{Dipartimento di Astronomia, Universit\`a degli studi di Padova,
Vicolo dell'Osservatorio 5, Padova, I-35122}
\altaffiltext{2}
{Dipartimento di Fisica, Universit\`a degli studi ``Roma Tre'',
Via della Vasca Navale 84, Roma, I-00146}
\begin{abstract}
Recent results have questioned the description of the QSO luminosity
function in terms of a pure luminosity evolution and call for a
luminosity dependent luminosity evolution.
Measurements of the QSO clustering amplitude and evolution allow
further distinguishing among the various physical scenarios proposed to
interpret the QSO phenomenon.  The general properties of the QSO
population would arise naturally if quasars are short-lived events
connected to a characteristic halo mass $\sim 5 \cdot 10^{12}$
M$_\odot$.  This is the typical mass of groups of galaxies in which
the interactions triggering the QSO activity preferentially take
place.
\end{abstract}
\vskip -10truemm
\section{Introduction}
QSOs have played an important role as cosmic probes of the young
universe: they are used as light beacons for absorption-line and
gravitational lensing studies, as markers of galaxy formation activity
and have guided astronomers in hunting up primeval galaxies.  Their very
existence at high redshifts is a challenge for cosmological models and
historically the evolution of the QSO population has been one of the
first evidences for an evolving universe.  
The nature of QSOs and the causes of their evolution, however, are far from
being fully understood. In the standard ``demographic'' approach 
the basic physical ideas are tested against the observed
shape and evolution of the QSO luminosity function (LF).
As usual, with the growth and improvement of the databases 
this type of interpretation has become significantly
more complex and we
will argue that to disentangle the physical evolutionary patterns of
QSOs additional information is needed, for example about their
clustering properties.
\section{The QSO Luminosity Function}
In the early nineties the general consensus among the astronomers was
that the shape of the QSO LF is a double power-law and its evolution,
at least up to $z\simeq 2.2$, can be satisfactorily represented by a
pure luminosity evolution (PLE) (Boyle {et al. 1988):
\begin{equation}
\Phi (M_B,z) = { {\Phi^\ast} \over { 10^{0.4[M_B-M_B(z)](\alpha +
1)} + 10^{0.4[M_B-M_B(z)](\beta + 1)} } }
\end{equation}
where $\alpha$ and
$\beta$ correspond to the bright-end and faint-end slopes of the
optical LF, respectively. With this parameterization the evolution of
the LF is uniquely specified by the redshift dependence of the break
magnitude, $M_B(z) = M_B^{\ast} - 2.5~k\log(1+z)$,
corresponding to a power-law evolution in the optical luminosity,
$L^{\ast}\propto (1+z)^k$. 
Such a result was essentially based on the AAT (Boyle et al. 1990) and
PG (Schmidt \& Green 1983) surveys with the addition of the faint data
in the Marano field (Zitelli et al. 1992). The relevant parameters
were estimated to be $\alpha = -3.9$, $\beta = -1.5$, $M_B^{\ast} =
-22.4$, $k_L = 3.45$ with a redshift cutoff $z_{\rm max} =1.9$ after
which the evolution stops.
\begin{figure}[t]
\plotfiddle{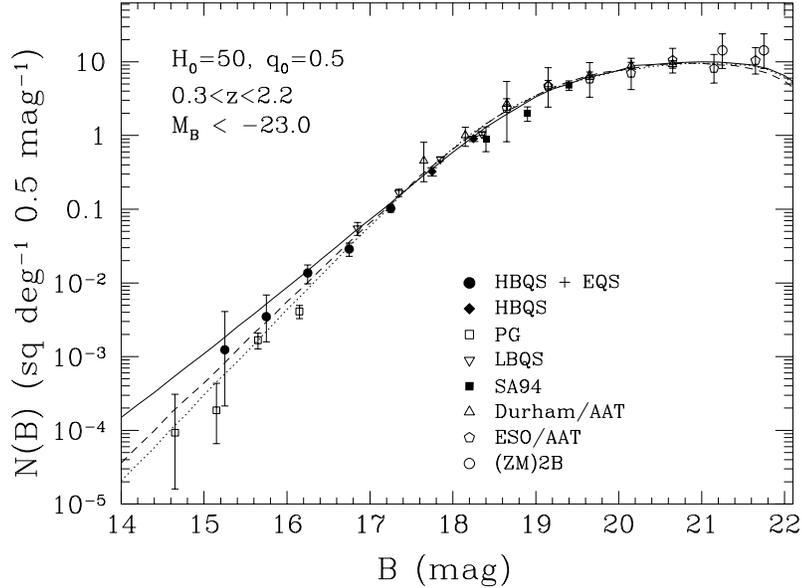}{200pt}{0}{40}{40}{-200}{-10}
\caption{
The QSO number counts for $0.3<z<2.2$. The continuous
line represents 
the best fit LDLE model, the dashed line represents our
best fit PLE model, while the dotted line represents the
best fit PLE model by Boyle {\it et al.} (1992).
}
\label{countfig}
\end{figure}
Then a new generation of bright surveys came: the LBQS (Hewett et al.
1995 and references therein), HBQS (Cristiani et al. 1995), EQS
(Goldschmidt et al. 1992), HES (Wisotzki et al. 1996). It turned out that 
not only the surface density of bright QSOs ($B \sim 15$) is
definitely higher than it was thought before, but also the new LF is
qualitatively different and cannot be reconciled with a PLE
(Fig.~\ref{lfnew}).
\begin{figure}
\plottwo{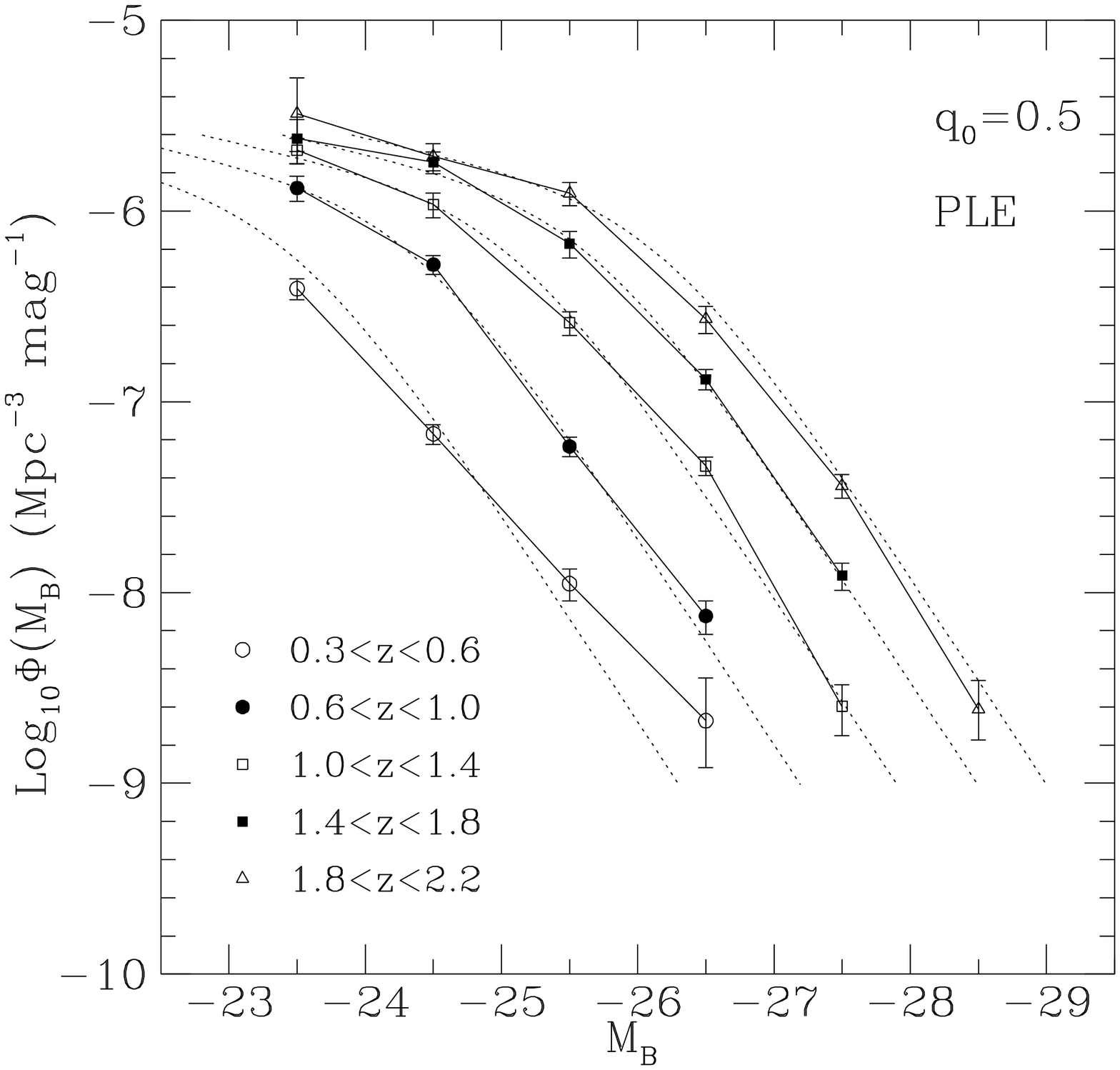}{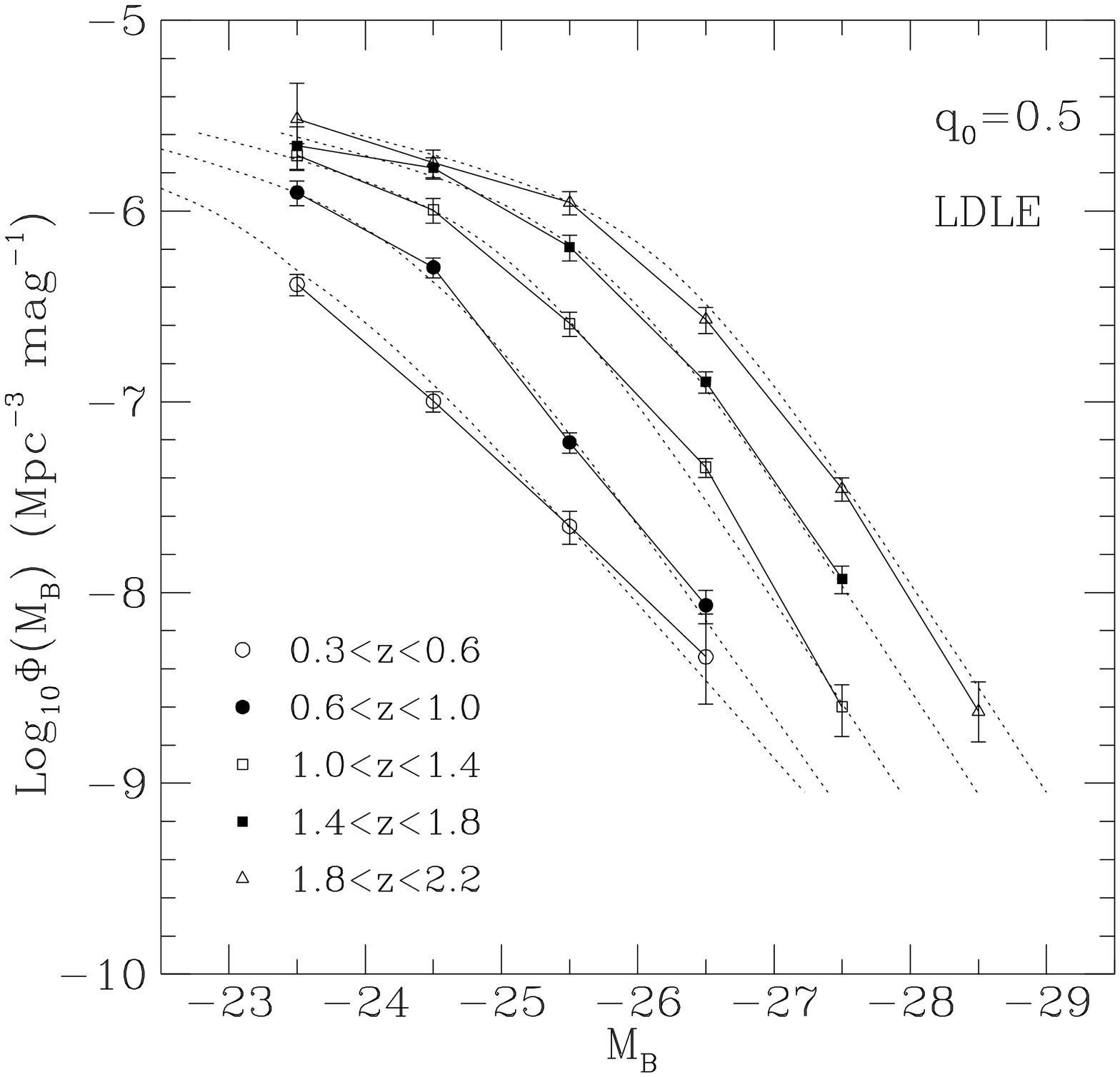}
\caption{ The QSO luminosity function. 
(Left) The dashed lines correspond to the best-fitting
Pure Luminosity Evolution with  $\alpha = -3.7$, $\beta = -1.4$,
$M_B^{\ast} = -22.3$, $k_L = 3.26$.
(Right) Fit with a 
Luminosity Dependent Luminosity Evolution model (see text).
The points connected with a continuous line represent the observations.
Error bars are based on Poisson statistic and
correspond to 68\% ($1 \sigma$) confidence intervals.
}
\label{lfnew}
\end{figure}
The flattening of the slope in the bright part of the low-z LF is apparent
and in the redshift bin $0.3 < z < 0.6$ the shape becomes consistent
with a single power-law (Goldschmidt \& Miller 1998). 
This behavior can be modeled with 
a luminosity dependent luminosity evolution (LDLE), in the form of an
evolution of the break magnitude:
$ M_B(z) = M_B^{\ast}(2) - 2.5k\log[(1+z)/3]$ where $M_B^{\ast}(2)$ is the
break magnitude at $z=2$ and
\begin{eqnarray}
\nonumber for~M_B \leq M_B(z)&:&~ k = k_1 + k_2 (M_B-M_B(z))e^{-z/{.40}} \\
\nonumber for~M_B > M_B(z)&:&~ k = k_1 . 
\end{eqnarray}
In this way the best-fitting parameters turn out to be
 $\alpha = -3.76$, $\beta = -1.45$, $M_B^{\ast}(2)=-26.3$,
$k_1=3.33$, $k_2=1.17$ (La Franca \& Cristiani 1997).
How robust is this result? Could this departure from PLE be due to
observational biases? \\
The new bright surveys have been carried out with different techniques:
UVx, slitless, and also cross-checked with other approaches, e.g. the
variability method (Hawkins \& V\`eron 1995), 
and the results are essentially matching.
Effects due to the photometric errors (the Bennet bias), the non-unique
SED of QSOs, the variability, the representation of the LF 
in discrete redshift bins are also accounted for 
(La Franca \& Cristiani 1997).
However, other biases could be lurking in: obscuration effects, either
intrinsic or intervening (Fall \& Pei 1993, Webster et al. 1995); 
effects due to the presence of the host
galaxy, especially for low-z, low-luminosity objects, although the
authors of various surveys state that this latter problem has been
considered and precautions taken accordingly.\\
In this respect, it will not be useless to compare the optical results
with radio and x-ray data.

Studies of radio QSOs show that there is an increase of the radio-loud
fraction among the bright ($M_B < -25$) QSOs, both in x-ray and
optically selected samples (La Franca et al. 1994, Hooper et al 1996). 
Possibly there is also an increase of the
radio-loud fraction at lower redshifts, but this result relies
entirely on the PG sample and should be considered with caution.
There are evidences in favor of a slower evolution of the optical LF
(OLF) of radio-loud QSOs, with a $2.7 < k_{RL} < 3.1$. 
Hook et al. (this conference, Fig.6) show that the space density
evolution is slower for more powerful radio sources.
In this way it
becomes possible to interpret the flattening of the bright part of the
OLF as a slower evolution of bright, radio-loud QSOs.

X-ray QSOs, on the other hand, seem to follow an evolution pretty
similar to the optically selected ones, with $k_x \simeq 3.2$ (Boyle
et al. 1994), which is not surprising, given the relationship between
the optical and X-ray luminosity of quasars (La Franca et al. 1995).
Recent deep and shallow ROSAT surveys provide evidences for departures
of the XLF from a PLE pattern (Hasinger 1998) and there might even be
an indication of flattening of the XLF at low-z and high luminosities
(see Fig. 1 of Hasinger 1998).

\section{Physical Considerations on the Luminosity Function}
At present there is a broad consensus that the QSO phenomenon is the
result of accretion onto super-massive black holes (BH) in the nuclei
of galaxies and a number of authors have linked the change in the QSO
activity to a corresponding change in fuel supply at the center of the
host galaxies (Cavaliere \& Szalay 1986; Wandel 1991; Small \&
Blandford 1992; Haehnelt \& Rees 1993; Yi 1996).  The epoch of QSO
formation is related to the time when the first deep potential wells
form in plausible variants of hierarchical cosmogonies and various
mechanisms for the evolution of the accretion modes have been explored
(see Haehnelt, Natarajan and Rees 1998, Cavaliere this conference) to
reproduce the observed luminosity function and its evolution.  
For example Cavaliere et al. (1997 and this conference) 
model the {\it rise and fall} of
the QSO LF as the effect of two components: the newly formed BH, which
are dominant at $z>3$ and the reactivated BH, which dominate at $z<3$.
The reactivation is triggered by interactions taking place
preferentially in groups of galaxies, of typical halo mass $\sim 5 \cdot
10^{12}$ M$_\odot$. 
In this scenario the flattening of the bright end of the LF at
low-z arises naturally due to the re-emergence of the first
component which is overwhelmed by the second at intermediate
redshifts.

In general, the agreement obtained by the various authors with the
observed LF within the supermassive BH paradigm and also outside
it (e.g. Boyle and Terlevich 1997) is remarkable.  The study of the
QSO clustering provides a powerful tool to unravel the riddle of the
interpretations.

\section{Clustering of QSOs}
QSO clustering was detected for the first time more than ten years ago
(Shaver 1984).  At the beginning the measurement was based on
inhomogeneous catalogs but soon it became possible to use the same
statistically-well-defined samples used for the LF works, although
with a significantly reduced number of QSOs.  
The two-point correlation function (TPCF, Peebles 1980) has been applied
as a standard approach to investigate the QSO clustering, but other
techniques have also been explored, such as the minimal spanning tree 
(Graham et al. 1995), fractal analysis (Andreani et al. 1991), 
counts in cells (Andreani et al. 1994).

In recent times complete samples totaling about 2000 QSOs 
have been used, providing a $4-5 \sigma$ detection of the clustering
on scales of the order of $6 h^{-1}$ comoving Mpc 
(Andreani \& Cristiani 1992, Mo \& Fang 1993, Croom \& Shanks 1996).  
The evolution of this clustering is
not clear.  An amplitude constant in comoving coordinates or
marginally decreasing with increasing redshift has been suggested, an
amplitude which appears to be consistent or slightly larger than what
is observed for present-day galaxies and definitely less than the
clustering of clusters.  Measurements of the quasar-galaxy correlation
indicate an amplitude about 4 times the galaxy-galaxy
clustering (Fisher et al. 1996).
\begin{figure}[t]
\plotfiddle{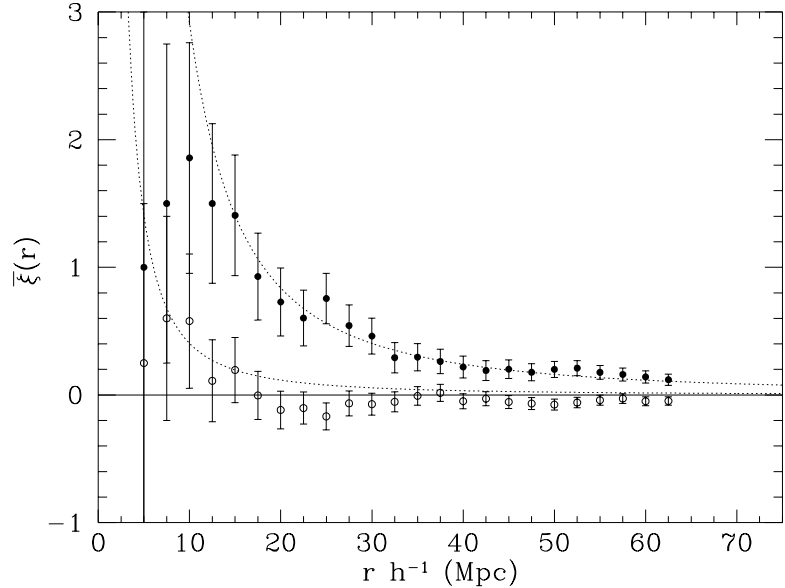}{200pt}{0}{40}{40}{-200}{-10}
\caption{
The integral correlation function $\bar\xi(r)$ - defined as $\bar\xi(r)
= {3\over r^3} \int_0^r x^2\xi(x)dx$ - for the quasars in the SGP
sample in two redshift ranges $0.3<z\leq 1.4$ (open circles), 
and $1.4< z\leq 2.2$ (filled circles).
\label{clustfig1}
}
\end{figure}
With all the caveats due to their exotic nature, QSOs display a
number of appealing properties when compared to galaxies as
cosmological probes of the intermediate-linear regime of clustering,
on distances $\geq 20$ Mpc: they have a rather flat $n(z)$, allow
defining samples which are in practice locally volume limited, their
point-like images are less prone to the surface-brightness biases
typical of galaxies and they sparse-sample the environment.
The sparse sampling is an advantage, in the sense that the
contribution of each QSO or QSO pair to the correlation functions in
practice turns out to be independent, making statistics easy, but is
also a nuisance, because the sensitivity to the clustering signal is
low.

In an attempt to improve the situation, while waiting for the 2dF QSO
redshift survey, we have carried out a survey in
the South Galactic Pole (SGP) over a {\it connected} area of 25 square
degrees down to $B_j = 20.5$. Stacked UKSTU
plates were used to select UVx candidates and the multi-fiber
spectrograph MEFOS at ESO to take spectra of them. The final sample is
made up of 388 QSOs with $0.3<z<2.2$ (La Franca et al 1998a). 

The TPCF analysis gives an amplitude $r_o = (6.2\pm1.6) ~h^{-1}$ Mpc, in
agreement with previous results. However, when the evolution of the
clustering with redshift is analyzed, evidence is found for an {\it
increase} of the clustering with increasing z
(La Franca et al. 1998b). The effect is small, at
a $2 \sigma$ level (Fig.~\ref{clustfig1}), but is interestingly
corroborated by other results in the literature.
\begin{figure}[t]
\plotfiddle{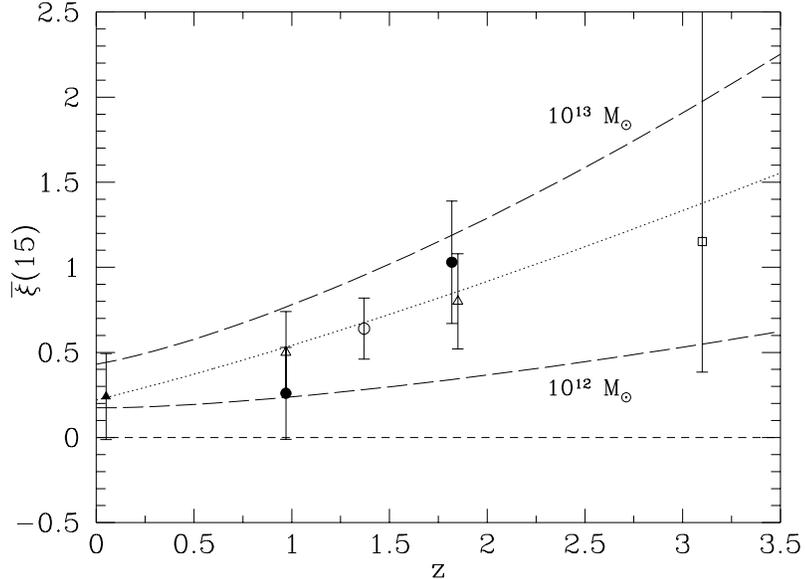}{200pt}{0}{40}{40}{-200}{-10}
\caption{
The amplitude of the $\bar\xi(15~h^{-1}$ Mpc) as a function of z.
Filled circles: the low- and high-z SGP subsamples; 
open circle: the SGP sample plus the 
Boyle et al. (1990), La Franca, Cristiani and Barbieri (1992), and Zitelli et
al. (1992) samples; open triangles: same as open circle but divided in
two redshift slices; filled triangle: low-z AGNs 
from Boyle and Mo (1993) and Georgantopoulos and Shanks (1994);
open square: the high-z sample from Kundi\'{c} 1997.
The dotted line is the $\epsilon=-2.5$ clustering evolution, 
the dashed lines are the $10^{12}$ and $10^{13}$ $M_{\odot}$ $h^{-1}$ minimum
halo masses clustering evolution according to the transient model of Matarrese
et al. (1997). 
}
\label{clustfig}
\end{figure} Boyle and Mo (1993) measured the clustering of low-z QSOs in the
EMSS and Georgantopoulos and Shanks (1994) used the IRAS point source catalog
to measure the clustering of Seyferts. Altogether a low value of the TPCF at
15 Mpc and z=0.05 is obtained, $\bar\xi = 0.24 \pm 0.25$. It may be argued
that these surveys tend to select low-luminosity objects and the comparison
with the SGP data could not be entirely significant, but an analysis on
restricted absolute magnitude slices of the SGP sample shows no correlation of
the clustering with the QSO absolute luminosity.  Besides, the data of the
Palomar Transit Grism Survey (Kundi\'{c} 1997, Stephens et al. 1997) allow
measuring the amplitude of the TPCF at redshifts higher than 2.7 and the
result, $r_o = (18\pm8) h^{-1}$ Mpc, suggests that the trend of increasing
clustering persists.

If we describe the evolving correlation function in a standard way: $\xi(r,z)
= ({r/{r_0}})^{-\gamma}(1+z)^{-(3-\gamma+\epsilon)}$, where $\epsilon$ is an
arbitrary (and not very physical) fitting parameter, we obtain $\epsilon = -
2.5\pm 1.0$, which is inconsistent with the $\epsilon \simeq 0.8$ observed for
faint galaxies at lower redshifts (Le F\`{e}vre et al. 1996, Carlberg et al.
1997, Villumsen et al. 1997). Great care should be exercised however when
carrying out this comparison. Are the faint lower-redshift galaxies
representative of the same population of galaxies for which recent
observations by Steidel et al (1998) show substantial clustering at $z \simeq
3.1$ ?  Are the Lyman-break galaxies progenitors of massive galaxies at the
present epoch or precursors of present day cluster galaxies (Governato et al.
1998)? Can a mass scale of the order $10^{12}~M_{\odot}$ be made consistent
with the observations of field galaxy clustering at all redshifts (Moscardini
et al. 1998) via an interplay between the clustering of mass (decreasing with
increasing redshift) and the bias (increasing with redshift)?

If we treat clustering according to the models in which quasars at
$z<3$ are associated with interactions, then we may relate their clustering
properties - following Matarrese et al. (1997) - to those of transient
objects, which is definitely different from the case of galaxies which,
depending on the physical scenario, can be associated with a merging or
object-conserving paradigm of long-lived objects. In this way the observed
clustering is the result of the convolution of the true clustering of the mass
with the bias and redshift distribution of the objects and differences in each
of these factors may lead to different results. If we think of QSOs as objects
sparsely sampling halos with $M > M_{\rm min}$ an increase of their clustering
is expected because they are sampling rarer and rarer overdensities with
increasing redshift. We may also ask what are the typical masses which allow
reproducing the observed clustering. As we can see from Fig.~\ref{clustfig} an
$M_{\rm min}= 10^{12} - 10^{13}~ M_{\odot}$ would provide the desired amount
of clustering and evolution, and would be also in pleasing agreement with the
predictions derived in Sect. 3 from the analysis of the QSO luminosity
function.


\begin{references} 
\reference Andreani, P., Cristiani, S., La Franca, F. 1991, \mnras,
253, 527
\reference Andreani, P., Cristiani, S. 1992, \apj, 398, L13
\reference Andreani, P., Cristiani, S., Lucchin, F., Matarrese, S., 
Moscardini, L. 1994, \apj, 430, 458
\reference Boyle B.J., Shanks T., Peterson B.A., 1988, MNRAS, 235, 935 
\reference Boyle, B. J., Fong, R., Shanks, T., Peterson, B.A. 1990, 
\mnras, 243, 1
\reference Boyle, B.J., Mo, H.J. 1993, \mnras, 260, 925
\reference Boyle B.J., Shanks T., Georgantopoulos I., Stewart G.C.,
Griffiths R.E. 1994, MNRAS, 271, 639 
\reference Boyle B.J., Terlevich, R.J. 1997, astro-ph/9710134
\reference Carlberg, R.G., Cowie, L.L., Songaila,
A., Hu, E.M. 1997, \apj, 484, 538
\reference Cavaliere, A., Szalay, A.S., 1986, \apj, 311, 589
\reference Cavaliere, A., Perri, M., Vittorini, V. 1997,
Mem.S.A.It., 68, 27
\reference Cristiani, S., La Franca, F., Andreani, P. et al. 1995, 
A\&AS, 112, 347
\reference Croom, S.M., Shanks, T. 1996, \mnras, 281, 893
\reference Fall, S.M., Pei, Y.C. 1993 \apj, 402, 479
\reference Fisher, K.B., Bahcall, J.N., Kirhakos, S., Schneider,
D. P. 1996 \apj, 468, 469
\reference Georgantopoulos, I., Shanks, T. 1994, \mnras, 271, 773
\reference Goldschmidt P., Miller L., La Franca F., Cristiani S., 1992, 
MNRAS, 256, 65p
\reference Goldschmidt P., Miller L. 1998, \mnras, 293, 107
\reference Governato, F., Baugh, C.M., Frenk, C.S., Cole, S., Lacey,
C.G., Quinn, T., Stadel, J. 1998, Nature in press
\reference Graham, M.J., Clowes, R.G., Campusano, L.E. 1995, \mnras,
275, 790
\reference Haehnelt, M.G., Rees, M.J. 1993, \mnras, 263, 168
\reference Haehnelt, M.G., Natarajan, P., Rees, M.J. 1998, MNRAS, in
press, astro-ph/9712259
\reference Hasinger, G. 1998, Astronomische Nachrichten in press,
astro-ph/9712342 
\reference Hawkins, M.R.S., V\`eron, P. 1995, MNRAS, 275, 1102
\reference Hewett P.C., Foltz C.B., Chaffee F.H., 1995, AJ, 109, 1498 
\reference Hooper, E.J., Impey, C.D., Foltz, C.B., Hewett, P.C.
1996, \apj, 473, 746
\reference Kundi\'{c}, T., 1997, \apj, 482, 631
\reference La Franca, F., Cristiani S., Barbieri, C. 1992, \aj, 103, 1062
\reference La Franca, F., Gregorini, L., Cristiani, S., De Ruiter,
H., Owen, F. 1994, \aj, 108, 1548
\reference La Franca, F, Franceschini, A., Cristiani, S., Vio, R. 
1995, A\&A  299, 19
\reference La Franca, F., Cristiani, S. 1997 \aj, 113, 1517; ERRATUM, 1998 \aj,
115, April issue 
\reference La Franca, F., Lissandrini, C., Cristiani, S., Miller, L.,
Hawkins, M.R.S., McGillivray, H.T., 1998a, in prep
\reference La Franca, F., Andreani, P., Cristiani, S. 1998b, \apj, in press
astro-ph/9711048
\reference  Le F\`{e}vre, O., Hudon, D., Lilly,
S.J., Crampton, D., Hammer, F., and Tresse, L. 1996, \apj, 461, 534
\reference Matarrese, S., Coles, P., Lucchin,
F., and Moscardini, L. 1997, \mnras, 286, 115
\reference Moscardini, L., Coles, P., Lucchin, F., 
Matarrese, S., 1998, MNRAS submitted, astro-ph/9712184
\reference Mo, H.J., Fang, L.Z. 1993, \apj, 410, 493
\reference Peebles, P.J.E. 1980, {\it The Large Scale Structure of the
Universe}, Princeton University Press, Princeton
\reference Schmidt, M., Green, R.F. 1983, \apj, 269, 352
\reference Shaver, P.A. 1984, A\&A, 136, L9
\reference Small, T.A., Blandford, R.D. 1992, MNRAS, 259, 725
\reference Steidel, C.C., Adelberger, K.L., Dickinson, M., Giavalisco,
M., Pettini, M., Kellogg, M 1998 \apj, 492, 428.
\reference Stephens, A.W., Schneider, D.P., Schmidt, M., Gunn, J.E., 
Weinberg, D.H., 1997, \aj, 114, 41
\reference Villumsen, J.V., Freudling, W., da Costa, L.N 1997, \apj, 481, 578
\reference Webster, R.L., Francis, P.J., Peterson, B.A., Drinkwater,
M.J., Masci, F.J. 1995, Nature 375, 469
\reference Wisotzki L., Kohler T., Groote D., Reimers D., 1996, 
A\&AS, 115, 227
\reference Yi, I. 1996, \apj, 473, 645
\reference Zitelli, V., Mignoli, M., Zamorani, G., Marano, B., 
Boyle, B.J. 1992, \mnras, 256, 349
\end{references}
\end{document}